\documentclass[prc,superscriptaddress,showpacs,amssymb,amsmath,
amsfonts,aps,twocolumn,secnumarabic,floatfix]{revtex4-1}
\usepackage{graphics}
\usepackage{epsfig}

\begin{document} 
\bibliographystyle{try} 

\topmargin -0.9cm 
 
 \title{Unitarity constraints on meson electroproduction at backward angles}

\newcommand*{\JLAB }{ Thomas Jefferson National Accelerator Facility, Newport News, Virginia 23606} 
\affiliation{\JLAB } 

\author{J.M.~Laget}
     \affiliation{\JLAB}

\date{\today} 

\begin{abstract} 

At large virtuality $Q^2$, the coupling to the $\rho$ meson production channels provides us with a natural explanation of the surprisingly large cross section of the $\omega$, as well as the $\pi^+$, meson electroproduction recently measured at backward angles, without destroying the good agreement between the Regge pole model and the data at the real photon point. Together with elastic re-scattering of the outgoing meson it also provides us with a way to explain why the node, that appears  at $u\sim -0.15$~GeV$^2$ at the real photon point, disappears at moderate virtuality $Q^2$. Predictions are given for the electroproduction of the $\pi^0$ meson.
\end{abstract} 
 
 
\maketitle 

\section{Introduction}

At first sight the recent measurement~\cite{Li19} of the cross section of the $\omega$ meson electro-production at backward angles is puzzling. On the one hand, it does not exhibit the node at $u\sim-0.15$ GeV$^2$ that exists in the real photo-production cross sections. On the other hand, it is much more larger (a factor around 50 to 100) than the expectation of the most obvious model: the nucleon Regge pole exchange~\cite{La00, La19} supplemented with the canonical nucleon dipole electromagnetic form factor~\cite{La04}.

\begin{figure}[hbt]
\begin{center}
\epsfig{file=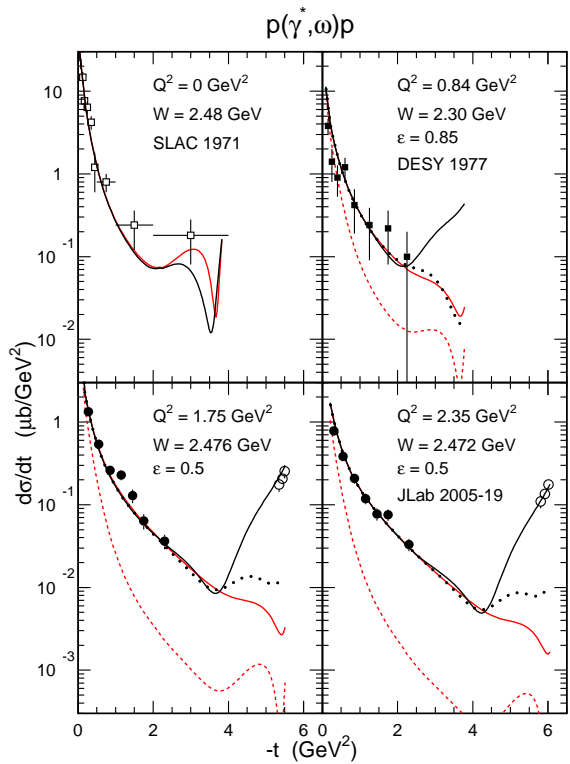,width=3.3in}
\caption[]{(Color on line) The cross section of the $p(\gamma^*,\omega )p$ reaction recorded at SLAC~\cite{Ba73}, top left, DESY~\cite{Jo77}, top right, and JLab~\cite{Li19,Mo05}, bottom left and right. The basic Regge Pole model~\cite{La19} with a non degenerated nucleon Regge trajectory  corresponds to the red curves: dotted line, when the canonical  pion electromagnetic form factor is used in the $t$-channel amplitudes; full line, when a $t$-dependent  cut-off mass is used~\cite{La04}. The black dotted line curves take into account the contribution of the $\rho^0 p$ scattering cut, while the black full line curves include also the contributions of the $\rho^+ n$ and  $\rho^{\pm} \Delta$ cuts. }
\label{dsdt_virtual}
\end{center}
\end{figure}

\begin{figure}[hbt]
\begin{center}
\epsfig{file=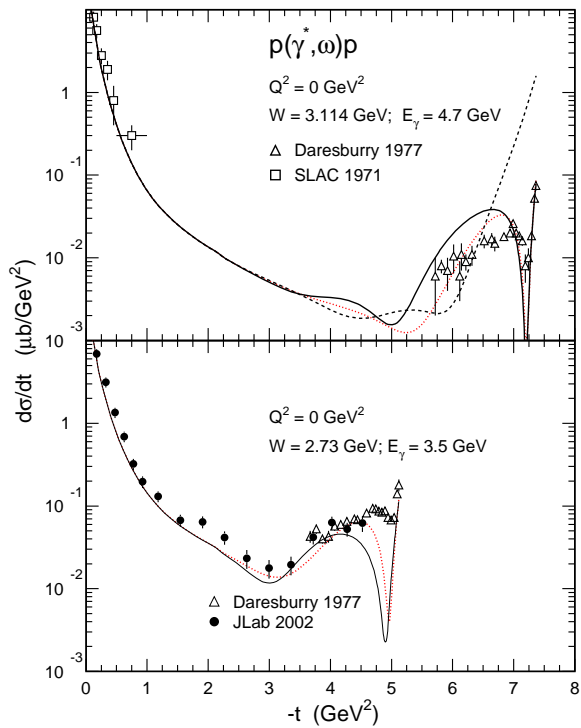,width=3.3in}
\caption[]{(Color on line) The cross section of the $p(\gamma,\omega)p$ reaction recorded at SLAC~\cite{Ba73}, Daresburry~\cite{Cl77} and JLab~\cite{Ba03}. The basic Regge Pole model~\cite{La19,La04} with a non degenerated Nucleon Regge trajectory corresponds to the red dotted line curves. The black dotted line curve uses the degenerated Nucleon trajectory. The black full line curves take also into account the contribution of the unitarity   scattering cuts. }
\label{dsdt_real}
\end{center}
\end{figure}

This resembles the photo- and electro-production of neutral pions at forward angles, that I succeeded to reproduce by a subtle, but straightforward, interplay between t-channel Regge poles and cuts~\cite{La11}. I have customized the same approach to the u-channel for omega production. The results are summarized in Figures~\ref{dsdt_virtual} and~\ref{dsdt_real}. The first deals with the evolution of the cross section with Q$^2$ at W around 2.47 GeV, in the full angular range. The second deals with the real photon point, at the energies where data exist in the full angular range. The physics is as follows:

At the real photon point, the most straightforward way to get a node in angular distributions at backward angles is to use a non degenerated Regge trajectory for the nucleon exchange in the u-channel. The red curves~\cite{La19} in Figure~\ref{dsdt_real} quantify this expectation which is supported by experiment, particularly at $E_{\gamma}=4.7$ GeV. However, when supplemented by the classical dipole nucleon electromagnetic form factor they badly miss the data~\cite{Li19}.  In Figure~\ref{dsdt_virtual}, the red dashed line curves are the predictions of this basic model when a constant cut off mass is used in the electric form factor of the pion that is exchanged in the $t$-channel. The red solid line curves are the predictions when a t dependent cut off mass is used, as explained in~\cite{La04}, to reproduce the JLab data~\cite{Mo05} at intermediate $-t$. At backward angles, the extrapolation of this $t$-channel contribution overwhelms the classical $u$-channel nucleon exchange contribution.  

\begin{figure}[hbt]
\begin{center}
\epsfig{file=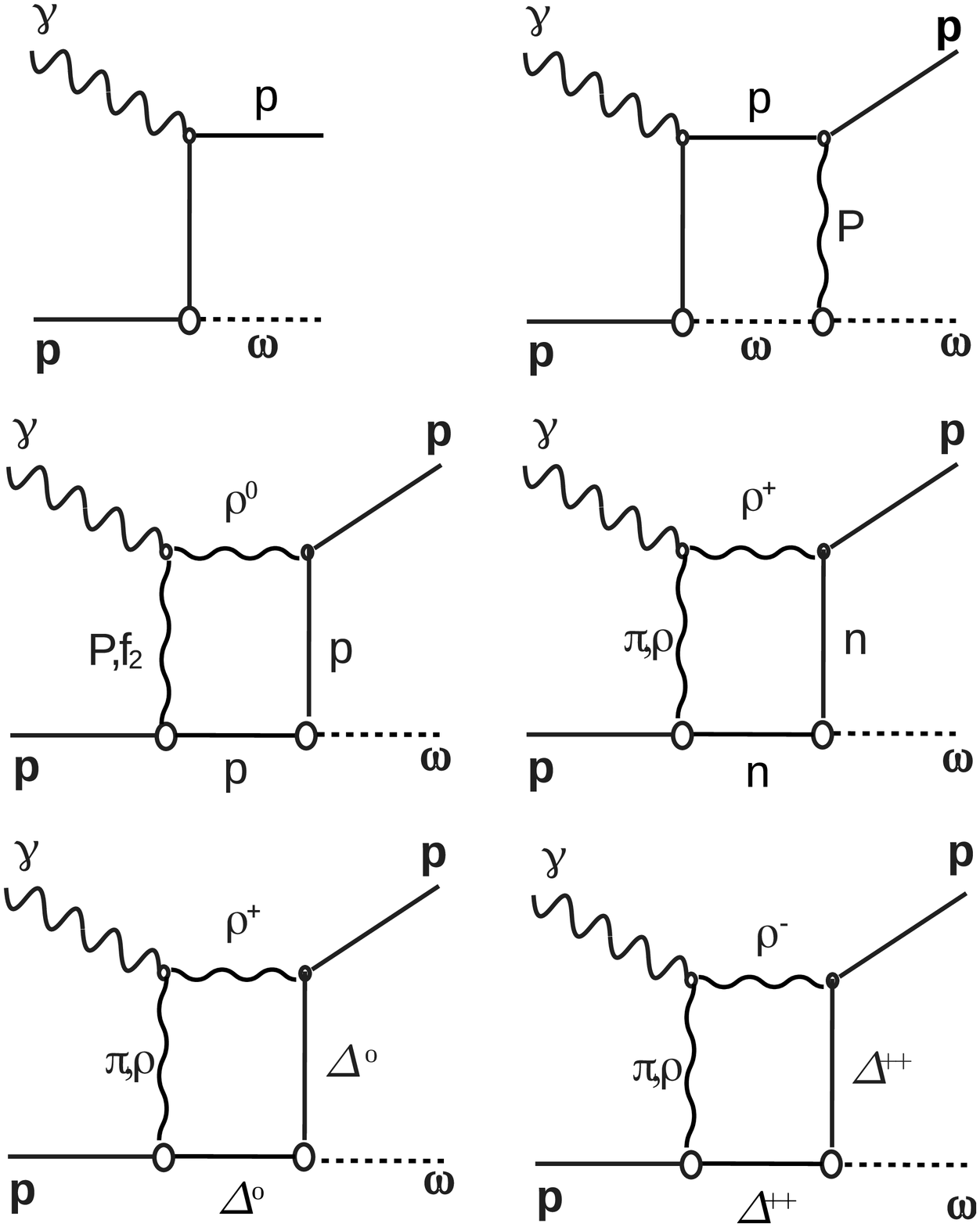,width=3.3in}
\caption[]{ Relevant graphs in the $p(\gamma^*,\omega)p$ reaction at backward angles. Top row: Nucleon Regge pole exchange (left) and elastic re-scattering cut (right). Middle row: $\rho^0 p$ (left) and $\rho^+ n$ (right) inelastic scattering cuts.  Bottom row: $\rho^+ \Delta^0$ (left) and $\rho^- \Delta^{++}$ (right) inelastic scattering cuts.}
\label{graphs}
\end{center}
\end{figure}

The other way to get a node is to add to the degenerated nucleon exchange amplitude (which does not exhibit a node) the elastic re-scattering cut (top right of Figure~\ref{graphs}), where the omega produced via nucleon exchange re-scatters on the nucleon. Since the omega nucleon elastic scattering amplitude is mostly absorptive, unitarity tells us that the corresponding cut interferes destructively with the nucleon pole: see equations 9-10-11 in~\cite{La10}. Also, the cut can be approximated by an effective Regge pole with a slope much smaller ($\alpha'$ about 0.2 GeV$^{-2}$) than the nucleon pole slope (about 0.98 GeV$^{-2}$): see equations 1-2-3 in~\cite{La11}. This interference leads to the black solid line curves in Figure~\ref{dsdt_real}. The black dashed line curve is the prediction of the nucleon degenerated pole only.

In the virtual photon sector, the nucleon exchange amplitude is driven by the canonical dipole electromagnetic form factor with a cut-off mass $\Lambda_N=0.7$ GeV$^2$. When this form factor is used in the elastic cut amplitude too, the data are badly underestimated at backward angles (red curves in the bottom part of Figure~\ref{dsdt_virtual}).

The electromagnetic cut off mass in the cut amplitude does not need to be the same as in the nucleon pole amplitude. This is a way to get rid of the node far from the photon point, but in order to get the measured cross section~\cite{Li19} at $Q^2=$ 2.35 GeV$^2$ one need to use an abnormally large cut off mass: $\Lambda_{cut}=$ 3.5 GeV$^2$.

Instead, this gap is easily explained by the contribution of the inelastic cuts.

The most obvious is the neutral rho meson production followed by the rho to omega transition via nucleon exchange (left part of the middle row in Figure~\ref{graphs}). Since the neutral rho production amplitude is dominated by the Pomeron exchange, this cut has the same structure, the same phase and the same Regge effective trajectory as the omega-nucleon elastic cut. It is therefore included implicitly in the fit of the elastic cut to reproduce the node at the photon point. Since the Q$^2$ dependency of the Pomeron exchange amplitude~\cite{La95} is much slower than the nucleon dipole form factor, the cross section is  significantly increased at backward angles (black dotted line curves in Figure~\ref{dsdt_virtual}). Also, the difference in the magnitude of the two components (pole and cut) prevents the formation of the node in their interference as $Q^2$ increases.

The rest of the gap is filled by the contribution of charged rho meson cuts (left part of middle row, and bottom row in Figure~\ref{graphs}) which brings the cross section close to experiment at the most backward angle. As noted in~\cite{La11} (see figure 8), the charged rho-nucleon production cross section represents only  one tenth of the neutral rho production cross section at the real photon point, but becomes comparable at Q$^2$=2.5 GeV$^2$. Also the charged rho-Delta production cross sections are comparable to the charged rho-N production cross section. When all these channels are added one gets the black solid line curves in Figure~\ref{dsdt_virtual}.

As we shall see later, a similar treatment reproduces also the cross section of the $\pi^+$ electro-production at backward angles recently measured at JLab~\cite{Pa18}, and permits to make predictions for the cross section of the $\pi^0$ electro-production. In both channels the $\Delta$ Regge pole exchange is also allowed (contrary to the $\omega$ production channel).

This paper quantifies this conjecture. The next Section deals with the modifications in the $u$-channel amplitudes ($t$-channel poles and cuts have not been modified and have been summarized in~\cite{La19}). Section~\ref{sec:real-photons} deals with the real photon sector, while Section~\ref{sec:virtual-photons} deals with the virtual photon sector. Predictions are given at energies higher ($W=\sqrt{s}\sim$~4~GeV) than available so far ($W\sim$~2.2~GeV).

\section{$u$-channel matrix elements}
\label{sec:mat-elem}

Let us start with the $p(\gamma,\omega)p$ reaction, where only the nucleon pole can be exchanged in the $u$-channel. The spatial part of the amplitude is as  follows:
\begin{eqnarray}
{\cal T}_N = i\frac{e\;g_{\omega}(1+\kappa_{\omega})}{2m}
  \left( \chi_f\left|\vec{\sigma}\cdot \vec{P_V}\times \vec{\epsilon_V} 
  \left[\mu_p\vec{\sigma}\cdot \vec{k}\times \vec{\epsilon}
  \right.\right.\right.
     \nonumber \\
   \left. \left. \left. 
  -i(2\vec{p_f}-\vec{k})\cdot\vec{\epsilon}-i(2\vec{p_i}+\vec{k})\cdot\vec{\epsilon}\;\frac{u-m^2}{s-m^2} \right]
                    \right| \chi_i \right)  						\nonumber \\
  \left\{ F_N(Q^2) \left(\frac{s}{s_0}\right)^{\alpha_N-\frac{1}{2}}                                                                                                              
    \frac{\pi \alpha'_Ne^{-i\pi (\alpha_N+\frac{1}{2})}}
           {\sin(\pi(\alpha_N+\frac{1}{2}))\Gamma (\alpha_N+\frac{1}{2})}
 \right.
 \nonumber \\
 \left. 
 -R _c F_c(Q^2) \left(\frac{s}{s_0}\right)^{\alpha_c-\frac{1}{2}} 
    \frac{G_c(u)\pi \alpha'_ce^{-i\pi (\alpha_N+\frac{1}{2})}}
           {\sin(\pi(\alpha_c+\frac{1}{2}))\Gamma (\alpha_c+\frac{1}{2})}
 \right\} \; \;\;
 \label{ome_amp}  	       
\end{eqnarray} 
where $e$ is the electric charge and $\mu_p$ is the magnetic moment of the proton, 
$\vec{k}$ is the momentum and $\vec{\epsilon}$ is the polarization of the incoming photon, while $\vec{P_V}$ is the momentum and $\vec{\epsilon_V}$ is the polarization of the emitted vector meson. The momenta of the incoming and the outgoing nucleons are respectively $\vec{p_i}$ and $\vec{p_f}$, while their energies are $E_i$ and $E_f$. The coupling constants of the $\omega$ meson with the nucleon are $g^2_{\omega}/4\pi=5.37$ and $\kappa_{\omega}=0$. They lie in between those that have been used in the study of the $\pi^0$ photo-production~\cite{La11} ($g^2_{\omega}/4\pi=4.9$) and the $\eta$ photo-production~\cite{La07} ($g^2_{\omega}/4\pi=6.44$) at forward angles. 

The first bracket exhibits the spin momentum structure of the amplitude. Only the magnetic coupling between the vector meson and the nucleon has been retained. While the convection part of the electromagnetic nucleon current has been taken into account too (it generates the Longitudinal part of the cross section). The $s$-channel nucleon pole is necessary to restore in a minimal way the gauge invariance of the convection current, while the magnetic current is gauge invariant by itself.

The first term in the curly bracket is the Regge propagator of the Nucleon pole amplitude. The energy scale is taken as $s_0=1$~GeV$^2$. The degenerated nucleon trajectory (with $\alpha_N(u) = -0.37 + \alpha'_N u$ and $\alpha'_N = 0.98$ GeV$^{-2}$) has been chosen. For convenience the Nucleon electromagnetic dipole form factor $F_N(Q^2)=1/(1+Q^2/0.7)^2$ is inserted here. The virtuality of the photon is $Q^2=-q^2=\vec{k}^2 -\omega^2$, where $\omega$ stands for the energy of the photon.

The second term is the Regge propagator which parameterizes the amplitude of the scattering cuts.  Since the elastic meson nucleon scattering amplitude and the $\rho^0$ meson electro-production amplitude are driven by the exchange of the Pomeron, which conserves helicity, the structure of the corresponding cuts is very similar and they can be combined. To a good approximation~\cite{Do02, colmart84}, the loop integral reduces to an effective Regge pole with the intercept and slope:
\begin{eqnarray}
\alpha_{c}(0)&=& \alpha_N(0)+ \alpha_{P}(0)-1\;\;\;\;\;\, = -0.37
\nonumber \\
\alpha'_c
&=& (\alpha'_N\times\alpha'_{P})/(\alpha'_N+\alpha'_{P})
= 0.2 \;\;\verb!GeV!^{-2}
\label{cut_traj}
\end{eqnarray}
where the intercept and the slope of the Pomeron Regge trajectory are respectively $\alpha_P(0)=1$ and $\alpha'_P= 0.25$ GeV$^{-2}$ (see~\cite{La19}). The form factor $G_c(u)$ takes into account the residual dependency upon $-u$ of the loop integral, while $R_c$ stands for the relative strength between the cut and the nucleon pole contribution. Both are fixed at the real photon point (Section~\ref{sec:real-photons}) in such a way to reproduce the results summarized in~\cite{La19}, which were based on the exchange on a non-degenerated Nucleon trajectory alone.

Eq.~\ref{ome_amp} exhibits explicitly the destructive interference between the Nucleon pole amplitude and the cut amplitude. Its comes from the fact that the Pomeron exchange amplitudes  are absorptive (imaginary) and that only the singular (imaginary) part of the rescattering loop integral has been retained (the demonstration parallels Eqs.~9, 10 and 11 of~\cite{La10}). It leads to the node in the cross section at backward angles.

For simplicity, the Regge phase of the cut is taken as the same as the phase of Nucleon pole. Since the intercept of their trajectory is the same, this choice does not affect the result at low $-u$, where these poles dominate.

The electromagnetic form factor $F_c(Q^2)$ of the cut does not need to be the same as the Nucleon form factor. If only the $\rho^0$N cut is considered, it rather follows the $Q^2$ dependency of the $\rho^0$ meson electro-production amplitude:
\begin{eqnarray}
F^{2g}_c(Q^2)&=& \frac{1}{(1+Q^2/(2\lambda_0^2+m_V^2))(1+Q^2/m_V^2)}
\nonumber \\
             &=& \frac{1}{(1+Q^2/6)(1+Q^2/0.6)}
\label{F2g}
\end{eqnarray}
according to the two gluon  exchange model of the Pomeron~\cite{La95}, with $\lambda_0^2=2.7$~GeV$^2$ and $m_V^2=0.6$~GeV$^2$.

\begin{figure}[hbt]
\begin{center}
\epsfig{file=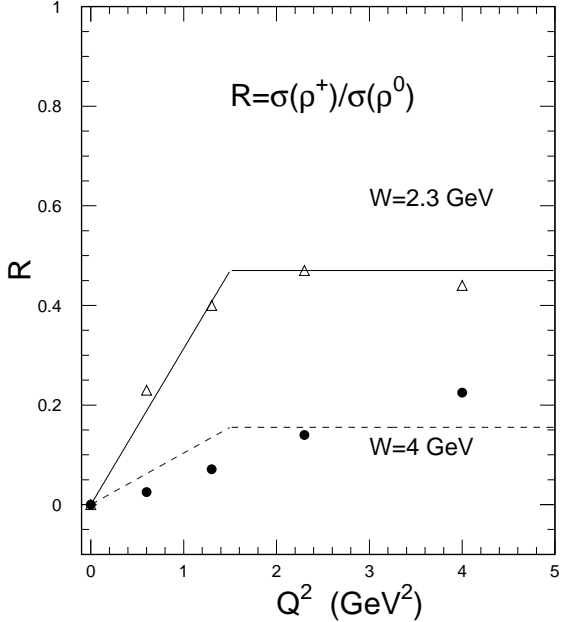,width=3.in}
\caption[]{ The ratio of the cross section of the $p(e,e'\rho^{\pm})N$ and $p(e,e'\rho^0)p$. The black circles and empty triangles correspond to the model~\cite{La00, La11}. The curves corresponds to the fit  Eq.~\ref{cross_section_ratio}.}
\label{Xsec_ratio}
\end{center}
\end{figure}

The contribution of the $\rho^+$n cut is negligible at the real photon point but becomes as important as the $\rho^0$p cut one in the virtual photon sector. We relate it to the ratio of the cross section of the $p(e,e'\rho^{\pm})N$ and $p(e,e'\rho^0)p$ reactions~\cite{La00, La11}, shown in Figure~\ref{Xsec_ratio}. We parameterize the ratio of the corresponding amplitudes as follows:
\begin{eqnarray} 
M(Q^2,W)&=& \frac{2.3}{W} \sqrt\frac{0.47Q^2}{1.5}
\;\;\; if\;Q^2\leq 1.5
\nonumber \\
&=& \frac{2.3}{W} \sqrt{0.47}
\;\;\;\;\;\;\; if\;Q^2 > 1.5
\label{cross_section_ratio}
\end{eqnarray} 

Since the charged $\rho$ production cross sections are driven by the exchange of the $\pi$ and the $\rho$ mesons while the neutral $\rho$ production cross section is driven by the exchange of the Pomeron, the ratio of the corresponding amplitudes i) behaves as W$^{-1}$ and ii) they must be added in quadrature. It is convenient to incorporate these features in the electromagnetic form factor of the cut, which becomes complex:
\begin{equation}
F_c(Q^2,W)= F^{2g}_c(Q^2)\left[1+i\;M(Q^2,W)\left(1
+R_{\Delta}
\frac{p_{\Delta}m_{\Delta}}{pm}\right)\right]
\label{Fc_full}
\end{equation}
The last term stands for the contribution of the charged $\rho\Delta$ cuts. As in~\cite{La11}, we assume that they have the same structure as the amplitude of the $\rho^+n$ cut, normalized by the ratio $R_{\Delta}$ of the relevant coupling constants and isospin coefficients. Its actual values will be determined in section~\ref{sec:virtual-photons}, and Appendix~\ref{AA}, for each channel. The ratio between the momentum $p_{\Delta}$ of the $\Delta$ and the momentum $p$ of the nucleon takes into account the difference between the phase-space available in the loop integral of the corresponding cuts.

Under the same assumptions, the Nucleon exchange amplitude in the $p(\gamma,\pi^+)n$ reaction is:
\begin{eqnarray}  
{\cal T}_N= -i\; e \mu_n g_{\pi} \sqrt{2}  \frac{\sqrt{(E_i+m)(E_f+m)}}{2m} 
\nonumber \\
 \left( \chi_f \left |
\vec{\sigma} \cdot \vec{k_{\gamma}} \times \vec{\epsilon}
 \;\;\vec{\sigma} \cdot \left[ 
 \frac{\vec{k_{\pi}}-\vec{p_i}}{\sqrt{m^2+ (\vec{k_{\pi}}-\vec{p_i})^2}+m} 
\right .\right .\right .    \nonumber \\ 
 \left . \left . \left . + \frac{\vec{p_i}}{E_i+m} \right]
\right | \chi_i \right )
\nonumber \\
  \left\{ F_N(Q^2) \left(\frac{s}{s_0}\right)^{\alpha_N-\frac{1}{2}}                                                                                                              
    \frac{\pi \alpha'_Ne^{-i\pi (\alpha_N+\frac{1}{2})}}
           {\sin(\pi(\alpha_N+\frac{1}{2}))\Gamma (\alpha_N+\frac{1}{2})}
 \right.
 \nonumber \\
 \left. 
 -R _c F_c(Q^2) \left(\frac{s}{s_0}\right)^{\alpha_c-\frac{1}{2}} 
    \frac{G_c(u)\pi \alpha'_ce^{-i\pi (\alpha_N+\frac{1}{2})}}
           {\sin(\pi(\alpha_c+\frac{1}{2}))\Gamma (\alpha_c+\frac{1}{2})}
 \right\} \; \;\;
\label{uN}
\end{eqnarray}
Where $(E_i, \vec{p_i})$ and $(E_f, \vec{p_f})$ are the four momenta of the target proton and the final neutron, respectively. Where $\vec{k_{\gamma}}$ is the momentum of the ingoing photon and $\vec{\epsilon}$ is its polarization. Where $(E_{\pi}, \vec{k_{\pi}})$ is the four momentum of the outgoing pion. The four momentum transfer in the $u$-channel is $u= (k_{\pi}-p_i)^2$. The magnetic moment of the neutron is $\mu_n=$ -1.91, and the pion nucleon coupling constant is $g^2_{\pi}/4\pi=$~14.5.

\begin{figure}[hbt]
\begin{center}
\epsfig{file=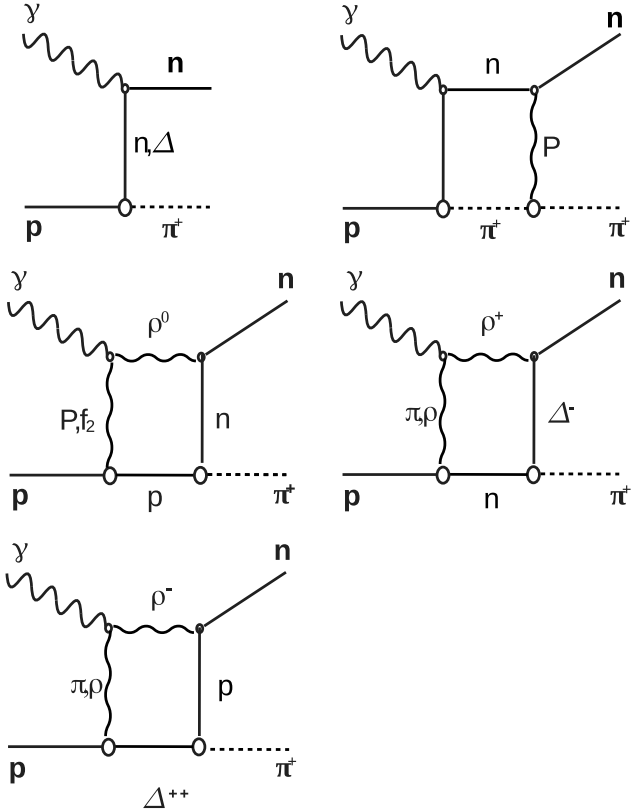,width=3.3in}
\caption[]{ Relevant graphs in the $p(\gamma^*,\pi^+)n$ reaction at backward angles. Top row: Nucleon and $\Delta$ Regge pole exchange (left) and elastic re-scattering cut (right). Middle row: $\rho^0 p$ (left) and $\rho^+ n$ (right) inelastic scattering cuts.  Bottom row: $\rho^-\Delta^{++}$ inelastic scattering cut.}
\label{graphs_piplus}
\end{center}
\end{figure}

The full relativistic expression of the $\pi NN$ coupling is used, while the lowest order of the $\gamma NN$ coupling is retained. Since the charge of the exchanged neutron is vanishing, there is no convection term. The electromagnetic form factor of the cut $F_c$ is complex and receives the contribution from the charged $\rho \Delta$ cuts that are  depicted in Figure~\ref{graphs_piplus} and which will be quantified in section~\ref{sec:virtual-photons}. 

The $\Delta$ exchange amplitude is the same as in~\cite{La10}:
\begin{eqnarray}  
T_{\Delta}&=& \frac{-1}{\sqrt{3}} e g_{\gamma \Delta} g_{\pi \Delta}  \frac{\sqrt{(E_i+m)(E_f+m)}}{2m} {\cal P}_{\Delta}^R(u)
\nonumber \\
&& \left( \lambda_f \left |
\vec{S}^{\dagger} \cdot \vec{k_{\gamma}} \times \vec{\epsilon}
 \;\;\vec{S} \cdot \vec{k_{\pi}} 
 \right | \lambda_i \right ) F_{\Delta}(Q^2)
 \label{DeltaN}
\end{eqnarray}
The magnetic coupling of the $\gamma N\Delta$ transition is $g_{\gamma \Delta}=$ 0.232~$(m_{\Delta}+m)/m_{\pi}$, and the pion Nucleon Delta coupling constant is $g_{\pi \Delta}=$~2.13/$m_{\pi}$ (see BL~\cite{BL77}). The electromagnetic form factor $F_{\Delta}(Q^2)$ is chosen identical to the Nucleon form factor $F_N(Q^2)$.
The  degenerated Regge propagator is:
\begin{eqnarray}  
{\cal P}_{\Delta}^R&=& \left (\frac{s}{s_{\circ}} 
\right)^{\alpha_{\Delta} -1.5}
 \alpha'_{\Delta}\;\; \Gamma (1.5-\alpha_{\Delta})
 \; e^{-i\pi(\alpha_{\Delta}-0.5)}
 \label{prop_Delta}
\end{eqnarray}
where the Delta trajectory is
$  \alpha_{\Delta}= 0.10 +\alpha'_{\Delta} \; u$,
with $\alpha'_{\Delta}=$~0.93~GeV$^{-2}$.

In the  $p(\gamma,\pi^0)p$ reaction the nucleon and $\Delta$ exchange amplitudes take the same form, with trivial changes of magnetic moment of the nucleon ($\mu_p= 2.78$ instead of $\mu_n=-1.91$), of the isospin coefficient of the $\pi NN$ vertex (1 instead of $\sqrt 2$) and the  $\pi \Delta N$ vertex ($\sqrt{2/3}$ instead of $\sqrt{1/3}$). 

As in the $\omega$ production channel the convection part of the nucleon electromagnetic  current should be added in Equation~\ref{uN}. However we do not take it into account in this study, since its contribution is negligible at the real photon point, where data exist, and since there is no data yet at backward angles in the virtual photon sector.

\begin{figure}[hbt]
\begin{center}
\epsfig{file=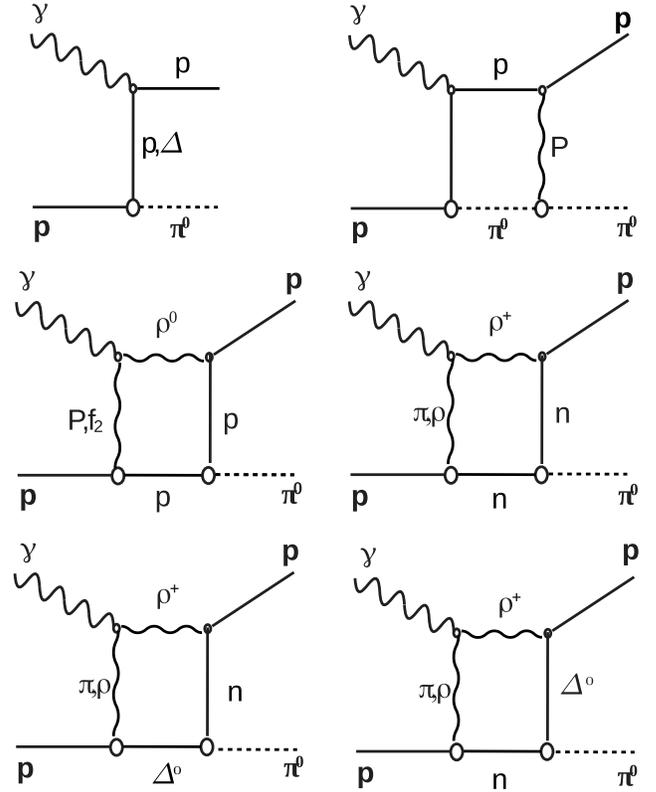,width=3.3in}

\caption[]{ Relevant graphs in the $p(\gamma^*,\pi^0)p$ reaction at backward angles. Top: Nucleon and $\Delta$ Regge pole exchange (left) and elastic re-scattering cut (right). Middle row: $\rho^0 p$ (left) and $\rho^+ n$ (right) inelastic scattering cuts. Bottom row: $\rho^+ \Delta^0$ (left) and $\rho^+ n$ (right) inelastic scattering cuts.}
\label{graphs_pizero}
\end{center}
\end{figure}

The charged $\rho N,\Delta$ cuts that contribute to the imaginary part of the cut form factor $F_c$ are depicted in Figure~\ref{graphs_pizero}. Their contribution will be quantified in section~\ref{sec:virtual-photons}.

\section{The real photon sector}
\label{sec:real-photons}
\subsection{$\omega$ photo-production}
\label{subsec:omega-real}

As already discussed in the Introduction, the interference between the proton degenerated Regge pole (black dotted line curve in Figure~\ref{dsdt_real}) and the elastic scattering (top right part in Figure~\ref{graphs}) and $\rho^0 p$ (middle left part in Figure~\ref{graphs}) unitary scattering cuts reproduces the reference model~\cite{La19} which is based on the exchange of  the non degenerated proton trajectory alone (red dotted line curves), as well as the experiments. To achieve this result the free parameters of the cut are chosen as:
\begin{eqnarray}
R_c &=& 4
\nonumber \\
G_c(Q^2) &=& e^{\lambda_c u}
\nonumber \\
\lambda_c &=& 0.7 \;\;\verb!GeV!^{-2}
\end{eqnarray}

\subsection{$\pi^0$ photo-production}

\begin{figure}[hbt]
\begin{center}
\epsfig{file=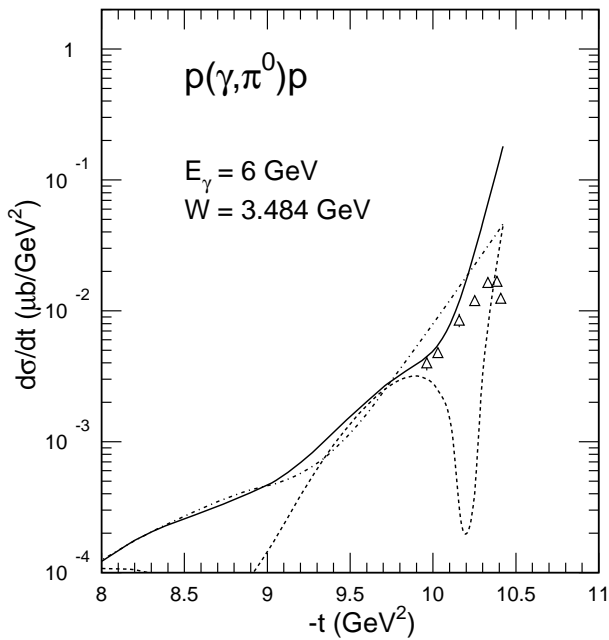,width=3.3in}

\caption[]{The cross section of the $p(\gamma,\pi^0)p$ reaction at $E_{\gamma}=6$~GeV. Experimental data~\cite{To69}. Dot-dashed line curves: $\Delta$ Regge pole exchange. Dashed line curves: nucleon Regge pole and unitary cuts. Full line curves: both contributions.}
\label{gam_pizero_6GeV}
\end{center}
\end{figure}

\begin{figure}[hbt]
\begin{center}
\epsfig{file=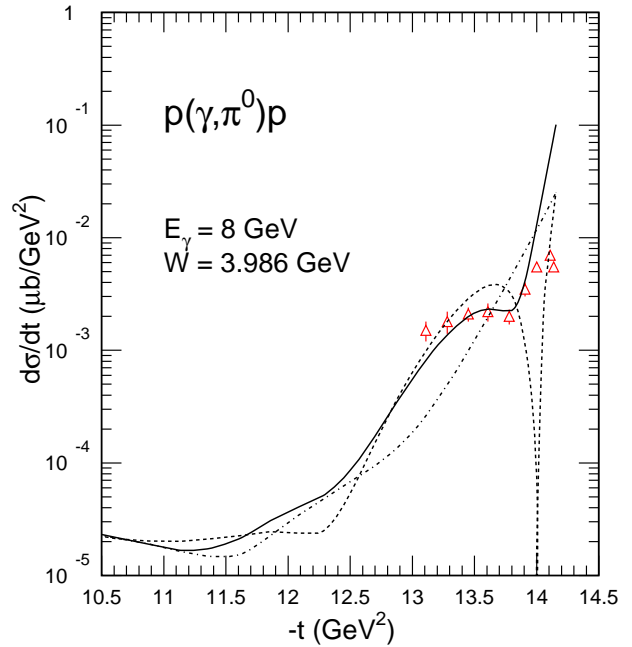,width=3.3in}

\caption[]{The cross section of the $p(\gamma,\pi^0)p$ reaction at $E_{\gamma}=8$~GeV. Data and curves: same meaning as in Figure \ref{gam_pizero_6GeV}.}
\label{gam_pizero_8GeV}
\end{center}
\end{figure}

Figures~\ref{gam_pizero_6GeV} and~\ref{gam_pizero_8GeV} compare the model to the  experimental data~\cite{To69} at backward angles that have been recorded at $E_{\gamma}=$~6 and 8~GeV.  The contribution of the $\Delta$ Regge pole exchange is comparable to the experiment and interferes with  the contribution of the Nucleon Regge pole exchange and the associated unitarity cuts. The parameters of the cuts, $R=4.25$ and $\lambda_c= 1.5$~GeV$^{-2}$, have been determined to best reproduce the experimental data at $E_{\gamma}=$~6~GeV. The canonical Blomquist-Laget~\cite{BL77} coupling constant  $g_{\pi\Delta}$ has been renormalized by $0.85$: this cosmetic liberty is marginal but  helps a slightly better reproduction of the data at both energies. Note that the u-channel amplitudes has been added to the tail of the t-channel amplitudes.

\begin{figure}[hbt]
\begin{center}
\epsfig{file=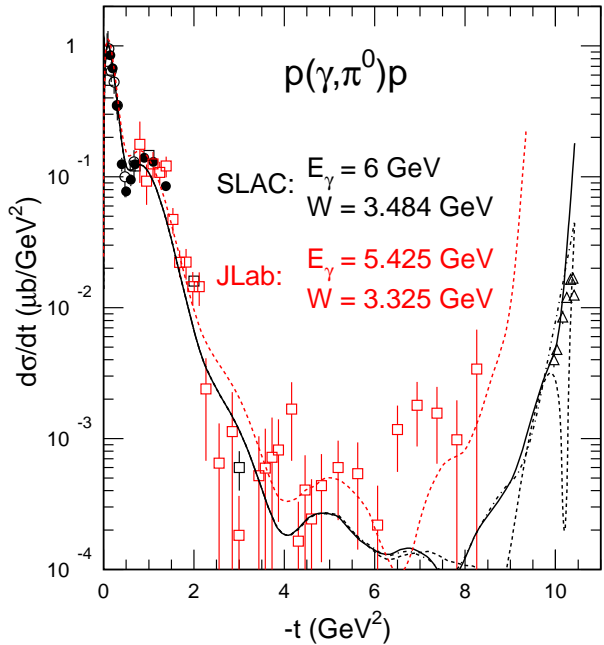,width=3.3in}

\caption[]{(Color online.) The cross section of the $p(\gamma,\pi^0)p$ reaction at $E_{\gamma}\sim6$~GeV, in the full angular range. The black line curves have the same meaning as in Figure \ref{gam_pizero_6GeV} at $E_{\gamma}=6$~GeV. The red dashed line curve is the prediction of the full model at $E_{\gamma}=5.425$~GeV. Open triangles~\cite{To69}, filled circles~\cite{An70}, empty squares~\cite{Ku18}.}
\label{gam_pizero_allt_6GeV}
\end{center}
\end{figure}

Figure~\ref{gam_pizero_allt_6GeV} compares the model to data~\cite{Ku18} which were recorded in the full angular range at the highest energy $E_{\gamma}=5.425$~GeV available at JLab. For completeness the SLAC data~\cite{To69, An70} recorded at $E_{\gamma}=6$~GeV at the most forward and backward angles are shown. The prediction of the model is given for the two energies, since the range in $-t$ is not the same at the most backward angles. 

\subsection{$\pi^+$ photo-production}

\begin{figure}[hbt]
\begin{center}
\epsfig{file=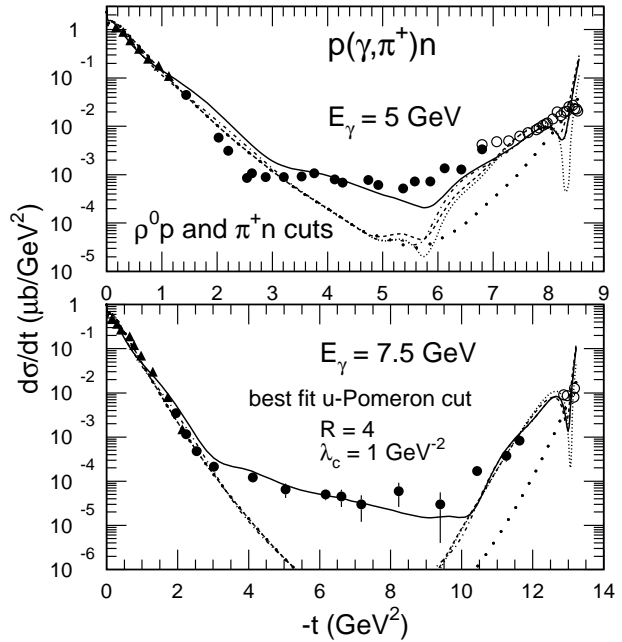,width=3.3in}

\caption[]{The cross section of the $p(\gamma,\pi^+)n$ reaction at $E_{\gamma}=5$~GeV and $E_{\gamma}=$~7.5~GeV. Triangles~\cite{Bo68}. Filled cicles~\cite{An76}. Empty circles~\cite{An69, An68}. The meaning of the each curve is given in the text.}
\label{gam_piplus_5GeV}
\end{center}
\end{figure}

Figure~\ref{gam_piplus_5GeV} compares the prediction of the model to the SLAC~\cite{An76, An69, An68, Bo68} experimental data that have been recorded at $E_{\gamma}=$~5~GeV and $E_{\gamma}=$~7.5~GeV. At backward angles, the contribution of the $\Delta$ pole exchange (double space dotted line curves) is smaller than the contribution of the nucleon exchange pole and elastic scattering cut (normal space dotted line curves). It cannot completely fill the node in the sum of both contributions (dash-dotted line curves). The parameters of the cut are:
\begin{eqnarray}
R_c &=& 4
\nonumber \\
\lambda_c &=& 1 \;\;\verb!GeV!^{-2}
\end{eqnarray} 
Note that the contribution of the $t$-channel poles has been retained in these curves. The contribution of the $t$-channel unitary $\pi$ elastic and $\rho^0$ inelastic rescattering cuts~\cite{La10} are included in the dash-dotted line curves and full line curves respectively.

\subsection{The link with previous approaches}

A first attempt has been carried out in the doctoral thesis of M. Guidal~\cite{Gui96}. While the details of the baryon Regge pole amplitudes have not been published elsewhere, the final results, for the $\pi^+$ photo-production channel, have been shown in Figure~1 of the  GLV paper~\cite{Gui97}. In the pseudo-scalar meson production channels, relativistic expressions of the electromagnetic vertices were used, and degenerated Baryon trajectories were used. Consistently with experiments, the corresponding cross section section did not exhibit a node, but were higher. In order to fit the experiment, the $u$-channel Regge amplitudes were renormalized by the factor:
\begin{eqnarray}
{\cal F}&=& \frac{2(\Lambda^2-m^2)^2}{\Lambda^4+(\Lambda^2-m^2)^2}
 \nonumber \\
\Lambda &=& 1.51 \;\verb!GeV!
\end{eqnarray}
which turns out to be ${\cal F} = 0.55$ for Nucleon exchange and ${\cal F} = 0.2$ for $\Delta$ exchange. Therefore the $\Delta$ exchange amplitude is more suppressed than the Nucleon exchange amplitude and plays a little role.

On the contrary a non degenerated Nucleon Regge trajectory has been used in the $\omega$ production channel, in order to  reproduce the node at backward angles. The large value of the coupling constant ($g_{\omega}= -15$ or $g_{\omega}^2/4\pi= 17.8$), consistent with the upper limit of the range determined in the analysis of nucleon-nucleon scattering, was renormalized by the factor ${\cal F}$. This is equivalent to a smaller value ($g_{\omega}= -8.21$ or $g_{\omega}^2/4\pi= 5.37$) in the lower limit of the range of accepted values.

The second attempt~\cite{La00, La19, La10} kept the degenerated trajectory of the $\Delta$, but used  the non degenerated trajectory of the nucleon in the $\omega$ meson as well as in the pseudo-scalar meson production channels, without the renormalization factor ${\cal F}$. The contribution of the Nucleon exchange becomes close to the data and the corresponding node in the pseudo-scalar meson production cross section is filled by the contribution of the $\Delta$ exchange and by the tail of the various rescattering cuts in the $t$ channel. In the $\omega$ production channel this model leads to results very similar (in shape and magnitude) to the Guidal's results, when the coupling constant  $g_{\omega}= -8.21$ or ($g_{\omega}^2/4\pi= 5.37$) is used~\footnote{Note1}.

Contrary to Guidal's work, this second approach relies on the lowest order non relativistic expression of the electromagnetic  currents. The good agreement between the two approaches, in the $\omega$ production channel, leads to infer that relativistic effects are not capital, at least in the limits of the model. 

In the present work, the degenerated Nucleon and $\Delta$ Regge trajectories are used in every channel. The interference between the poles and the elastic rescattering and $\rho^0 N$ unitarity cuts not only reproduces the node in the cross sections at backward angles, but also reduces the contribution of the Nucleon pole and brings the cross section down to the experiment.

At the real photon point, this model leads by construction to results  almost identical to the results that has been summarized in the review~\cite{La19}, of which the conclusions remain the same. The two components (Regge poles and cuts) of the amplitude at backward angles provides us with a different reference point  which  allows more freedom when extrapolating to the virtual photon sector.

For the sake of completeness Table~\ref{table_elastic_cut} collects the constants of the cuts, as well as the $\Delta$ pole, that are used in the three channels at the real photon point. They are comparable, and the slight differences may come from the particular relative importance of the Nucleon and $\Delta$ poles in each channel. They are similar to the values of $R_c= 3.7$ and $\lambda_c=2$~Gev$^{-2}$ that have been used in the similar study~\cite{La11} of the node which occurs at forward angle in $\pi^0$ photo-production.

\begin{center}
\begin{table}[htb]
\begin{tabular} {|c|c|c|c|c|}\hline
\multicolumn{5}{|c|}{\bf Unitarity cut constants}\\ \hline
\multicolumn{1}{|c|}{Channel} & {$R_c$} & {$\lambda_c$} & {$R_{\Delta}$} & {$g_{\pi \Delta}$} \\ \hline \hline
$\omega$ & 4    & 0.7 & 1.9 & 0 \\
$\pi^0$  & 4.25 & 1.5 & 1.5 & 0.85 BL \\
$\pi^+$  & 4    & 1   & 1.5 & BL \\ \hline
\end {tabular}
\caption[] {The parameters of the unitarity cuts. The unit for $\lambda_c$ is GeV$^{-2}$. The fourth column gives the values of the coefficients of the charged $\rho N,\;\Delta$ cuts that are determined in the next section. For convenience the last column gives also the values of the $\pi N\Delta$ coupling constant (BL means the canonical value~\cite{BL77}).}
\label{table_elastic_cut}
\end{table}
\end{center}

\section{The virtual photon sector}
\label{sec:virtual-photons}

\subsection{$\pi^+$ electro-production}

\begin{figure}[hbt]
\begin{center}
\epsfig{file=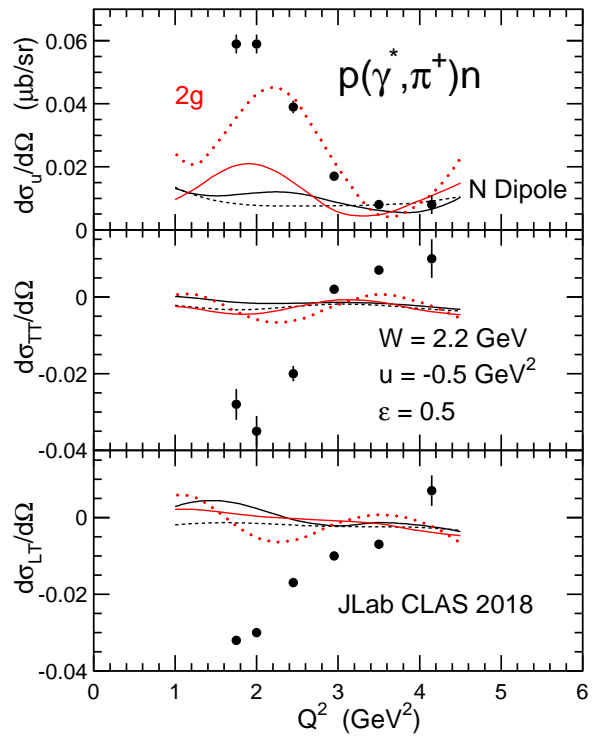,width=3.3in}

\caption[]{(Color online.) The cross section of the $p(\gamma^*,\pi^+)n$ reaction at $W=2.2$~GeV and $-u= 0.5$~GeV$^2$. Top: unpolarized cross section $\sigma_u=\sigma_T + \epsilon \sigma_L$. Middle: Transverse-Tranverse interference cross section. Bottom: Longitudinal-Transverse interference cross section. Filled circles~\cite{Pa18}. See text for the meaning of the curves.}
\label{gam_star_piplus_JLab_u0p5}
\end{center}
\end{figure}

Figure~\ref{gam_star_piplus_JLab_u0p5} shows the evolution with $Q^2$ of the cross section of the $p(\gamma^*,\pi^+)n$ reaction recently recorded~\cite{Pa18} at $-u=0.5$~GeV$^2$. The black curves use the canonical dipole form factor of the nucleon $F_N(Q^2)$ in both the pole amplitudes and the $\rho^0 p$ $u$-channel cut amplitude. The dashed line curves take into account the u-channel poles and cut (Equation~\ref{uN}), as well as the tails of the $t$-channel poles (negligible), while the full line curves include also the tail of  $t$-channel unitarity cuts~\cite{La19, La10}. The red full line curves use the two-gluons inspired form factor $F_c=F^{2g}_c$, Equation~\ref{F2g}, for the $\rho^0 n$  u-channel cut amplitude. The red dotted line curves take also into account the contribution of the  charged $\rho$~N, $\Delta$ cuts, with 
\begin{equation} 
R_{\Delta}=1.5
\end{equation}
in the expression of the complex form factor of the cuts $F_c(Q^2,W)$, Equation~\ref{Fc_full}. The determination of $R_{\Delta}$ is detailed in Appendix~\ref{AA}, while the definitions of the cross sections, $\sigma_u$, $\sigma_{TT}$ and $\sigma_{LT}$, as well as the polarization $\epsilon$ of the virtual photon are reminded in Appendix~\ref{AB}.

At large $Q^2$, the contribution of the $u$-channel cuts fills the gap between the large unpolarized experimental cross section $\sigma_u$ and the predictions of the poles alone. The disagreement at lower $Q^2$ may come from the fact that the $c.m.$ energy $W=2.2$~GeV is not high enough to get rid of the tail of the contributions of the $t$-channel poles and cuts. At such high values of $-t$, the resulting interference with $u$-channel contributions is not fully under control. 

\begin{figure}[hbt]
\begin{center}
\epsfig{file=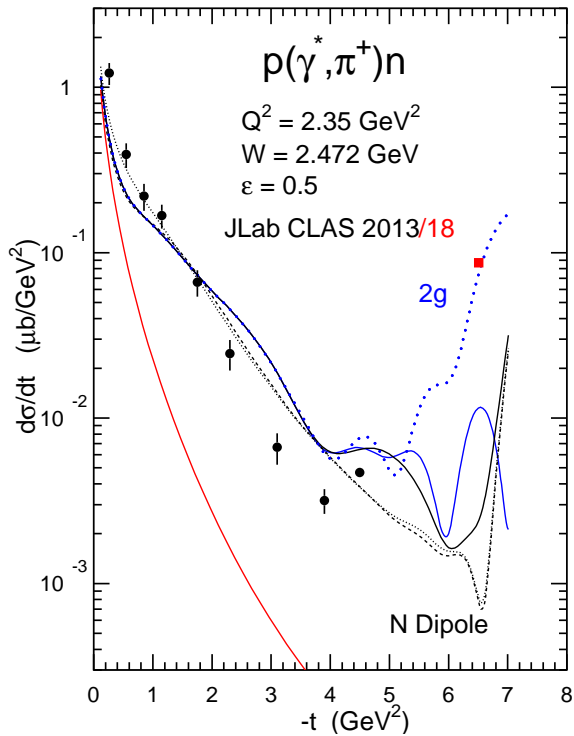,width=3.3in}

\caption[]{(Color online) The cross section of the $p(\gamma^*,\pi^+)n$ reaction at $W=2.45$~GeV and $Q^2=2.35$~GeV$^2$. Filled circles~\cite{Pa13}; filled red squares~\cite{Pa18}, at $W=2.2$~GeV$^2$. See text for the meaning of the curves.}
\label{gam_star_piplus_JLab_fullrange}
\end{center}
\end{figure}

This is illustrated in Figure~\ref{gam_star_piplus_JLab_fullrange} which compares the model to the data~\cite{Pa13, Pa18} in the entire angular range. The red full line curve takes only into account the $\pi$ and $\rho$ $t$-channel poles with $t$-independent electromagnetic form factor~\cite{Va98}. The black dashed line curve uses the $t$-dependent pion form factor which has been proposed in~\cite{La04}, and includes the $u$-channel poles cuts under the assumption $F_c(Q^2)=F_N(Q^2)$. The black dotted line curve takes also into account the contribution of the $t$-channel $\pi n$ elastic cut, while the black full line curve include also the contribution of the $t$-channel  $\rho p$ inelastic cuts~\cite{La10}. The blue full line curve uses $F_c(Q^2)=F_c^{2g}(Q^2)$, while the blue dotted line curve includes also the charged $\rho$-N,~$\Delta$ contributions in the $u$-channel.

\begin{figure}[hbt]
\begin{center}
\epsfig{file=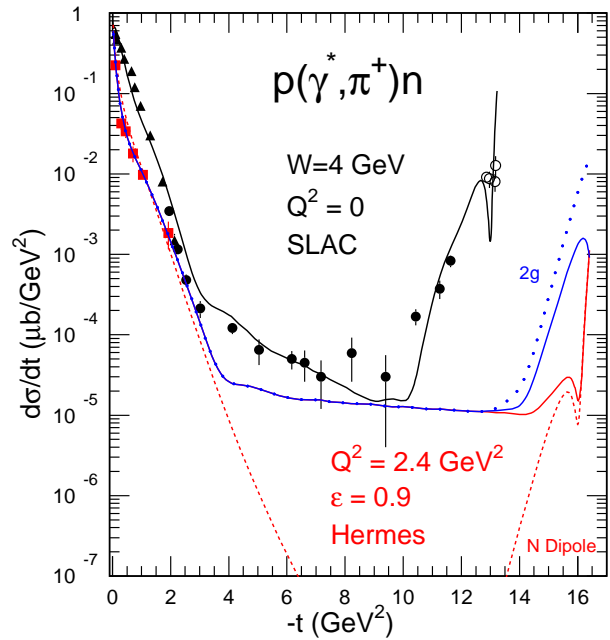,width=3.3in}

\caption[]{(Color online.) The cross section of the $p(\gamma^*,\pi^+)n$ reaction at $W=4$~GeV and $Q^2=2.4$~GeV$^2$. The black curve and experimental points are the same as in Figure~\ref{gam_piplus_5GeV} ($E_{\gamma} = 7.5$~GeV). Red filled squares~\cite{Ai08}. See text for the meaning of the curves.}
\label{dsdt_hermes_piplus}
\end{center}
\end{figure}

Such a mixing of the amplitudes becomes less severe (and even disappears) when the available energy increases. For instance in Figure~\ref{dsdt_hermes_piplus}, which displays the cross sections at $W=4$~GeV, the contributions of the $t$-channel poles~\cite{Gui97,Va98}, of the $t$-channel unitarity cuts~\cite{La10} and of $u$-channel poles and cuts are well separated at small, intermediate and large values of $t$ respectively.  At the real photon point, the  data and the curve are the same as in Figure~\ref{gam_piplus_5GeV} and are shown for reference. In the virtual photon sector,  $Q^2=2.4$~GeV$^2$, the forward angle data have been recorded at Hermes~\cite{Ai08}, while the full angular distribution will be accessible at JLab12~\cite{Li20} and the planned Electron Ion Collider (EIC)~\cite{EICxx}. The red dashed line curve includes the contributions of the $t$-channel poles (with the $t$~dependent Electromagnetic form factor of the $\pi$~\cite{La04}) as well as the $u$-channel poles and cuts with $F_c(Q^2)=F_N(Q^2)$. The red solid line curve includes also the contribution of $t$-channel unitarity cuts~\cite{La10}. Up to $-t=12$~GeV$^{-2}$ these curves are identical to the corresponding curves in Figure~25 of~\cite{La19}. By construction they are almost the same at the highest $-t$ values. The blue solid line curve uses the two gluon inspired form factor ($F_c(Q^2)=F^{2g}_c(Q^2)$, Equation~\ref{F2g}), while the blue dotted curve  also includes the charged $\rho$-N,~$\Delta$ $u$-channel cut contributions.

\begin{figure}[hbt]
\begin{center}
\epsfig{file=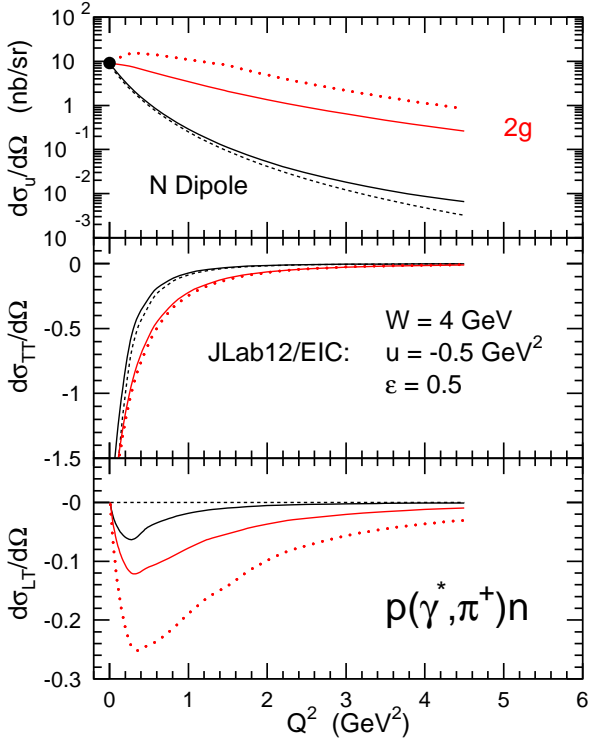,width=3.3in}

\caption[]{(Color online.) The cross section of the $p(\gamma^*,\pi^+)n$ reaction at $W=4$~GeV and $u=-0.5$~GeV$^2$. The black dot at the real photon point is the  experimental datum~\cite{An69, An68}. See text for the meaning of the curves.}
\label{u0p5_hermes_piplus}
\end{center}
\end{figure}

Figure~\ref{u0p5_hermes_piplus} shows the evolution of the cross section with the photon virtuality $Q^2$ at fixed $u=-0.5$~GeV$^2$ and $W=4$~GeV, an energy higher than in Figure~\ref{gam_star_piplus_JLab_u0p5}.  The black dotted line curves use the nucleon electromagnetic form factor for the nucleon pole as well as for the u-channel cuts ($F_c(Q^2)= F_N(Q^2$)), while the black solid line curves  include also the tail of the $t$-channel poles and cuts. The red solid line curves use $F_c(Q^2)=F_c^{2g}(Q^2)$, while the red dotted line curves include also the contribution of the charged $\rho$-N, ~$\Delta$ cuts. The cross sections are more regular at low $Q^2$, and exhibit a non trivial behavior when $Q^2$ increases: the full unpolarized cross section (dotted line curve) starts to increase, reaches a maximum around $Q^2=1$~GeV$^2$  and decreases above more slowly that the nucleon electromagnetic form factor. This is the reflection of the behavior  of the neutral and charged $\rho$ meson electro-production cross sections that enter the u-channel unitarity cut amplitudes. The planned measurement~\cite{Li20} at JLab12 will be useful  to further test this conjecture.

Note that the model is not expected to predict correctly the longitudinal cross section $\sigma_L$, at backward angles.  Since the electromagnetic current of the neutron pole is purely transverse, at the lowest order retained in the present work, the longitudinal amplitude is driven by the tails of the $t$-channel poles and cuts~\cite{La10} only. So the corresponding longitudinal and interference cross sections shown in Figures~\ref{gam_star_piplus_JLab_u0p5} and~\ref{u0p5_hermes_piplus} should be taken with care. A more accurate evaluation requires a complete treatment of the $u$-channel cut integrals, which is beyond the scope the present study.

\subsection{$\pi^0$ electro-production}

\begin{figure}[hbt]
\begin{center}
\epsfig{file=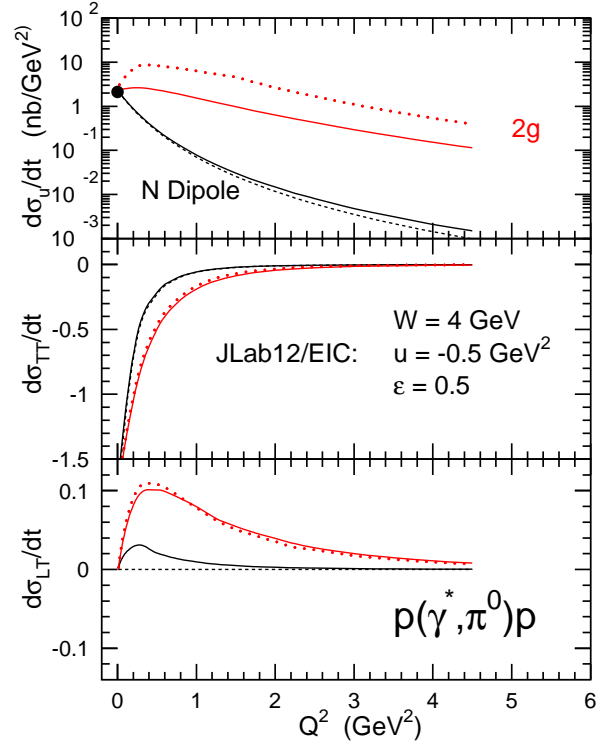,width=3.3in}

\caption[]{(Color online.) The cross section of the $p(\gamma^*,\pi^0)p$ reaction at $W=4$~GeV and $u=-0.5$~GeV$^2$. The black dot at the real photon point is the  experimental datum~\cite{To69}. The meaning of the curves is the same as in Figure~\ref{u0p5_hermes_piplus}.}
\label{u0p5_hermes_pizero}
\end{center}
\end{figure}

In this channel, there is no experimental data in the energy range considered in the present study. Figure~\ref{u0p5_hermes_pizero} shows the evolution of the cross sections with the photon virtuality $Q^2$ at an energy $W=4$~GeV reachable at JLab12~\cite{Li20}. The black line curves use the nucleon electromagnetic form factor $F_N(Q^2)$ for both the nucleon pole and the cut. The red solid line curves use the two gluon inspired form factor of the cuts, $F_c(Q^2)=F^{2g}_c(Q^2)$,  while the red dotted line curve takes into account the contribution of the charged $\rho \Delta$ cuts with the coefficient
\begin{equation} 
R_{\Delta}=1.5
\end{equation}
in the expression of the complex form factor of the cuts $F_c(Q^2,W)$, Equation~\ref{Fc_full}.
 
\begin{figure}[hbt]
\begin{center}
\epsfig{file=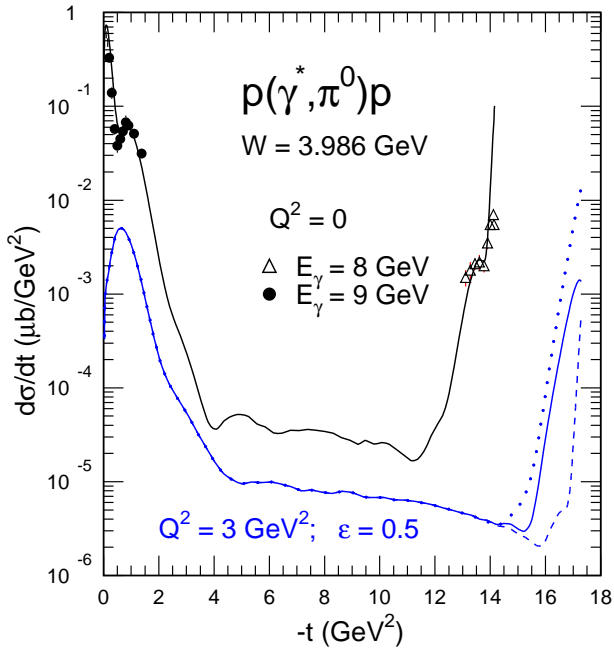,width=3.3in}

\caption[]{(Color online.) The cross section of the $p(\gamma^*,\pi^0)p$ reaction at $W=4$~GeV and $Q^2=3$~GeV$^2$ (blue curves). The black curve and experimental points at $E_{\gamma} = 8$~GeV are the same as in Figure~\ref{gam_pizero_8GeV}. The experimental points at $E_{\gamma}= 9$~GeV are from~\cite{An71}. The meaning of the blue curves is the same as in Figure~\ref{dsdt_hermes_piplus}.}
\label{dsdt_hermes_pizero}
\end{center}
\end{figure}

Figure~\ref{dsdt_hermes_pizero} shows the complete angular distribution at fixed $W=4$~GeV and $Q^2=3$~GeV$^2$. The real photons data and predictions are given for reference. Again, the use of the two gluons inspired form factor (blue solid line curve) and the contribution of the  charged $\rho \Delta$ cuts (blue dotted line curve) significantly increase the cross section at the most backward angles. 

\subsection{$\omega$ electro-production}

The comparison of the model with the JLab6 data~\cite{Li19} has been already presented in Figure~\ref{dsdt_virtual} and discussed in the Introduction. The red curves use the canonical dipole Nucleon form factor $F_N(Q^2)$ in both the nucleon pole and the u-channel cuts. The black dotted line curves use the two gluon inspired form factor $F^{2g}_c(Q^2)$ in the $\rho^0 p$ cut, while the black solid line curve takes also into account the contribution of the charged $\rho N$ and $\rho \Delta$ u-channel cuts, with the coefficient
\begin{equation} 
R_{\Delta}=1.9
\end{equation}
in the expression of the complex form factor of the cuts $F_c(Q^2,W)$, Equation~\ref{Fc_full}.

\begin{figure}[hbt]
\begin{center}
\epsfig{file=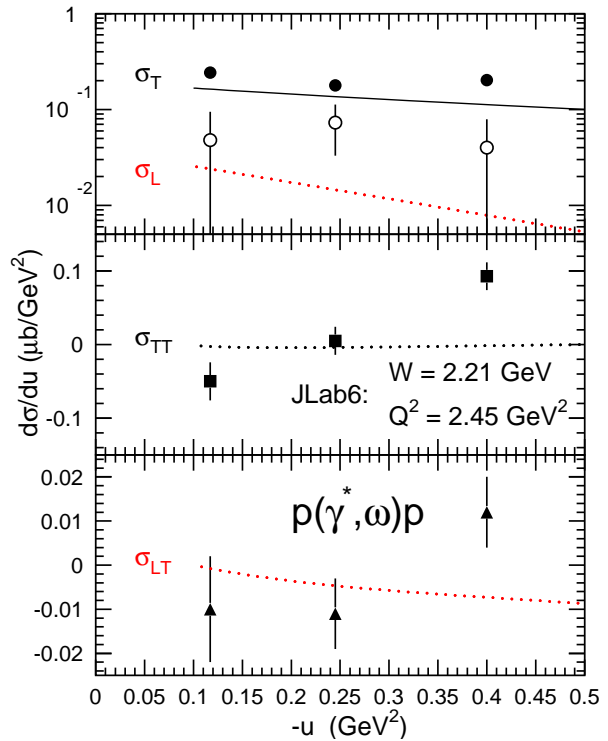,width=3.3in}

\caption[]{(Color online.) The separated cross sections of the $p(\gamma^*,\omega)p$ reaction at $W=2.21$~GeV and $Q^2=2.45$~GeV$^2$. In  the top window, the black solid line theoretical curve and the black filled  circle experimental points correspond to the Transverse cross section, while the dotted line curve  and the empty circle correspond to the Longitudinal cross section. The experimental data come from~\cite{Li19}.}
\label{dsdu_Jlab-separated}
\end{center}
\end{figure}

The model reproduces also the separated Transverse and Longitudinal cross sections recorded at $W=2.21$~GeV and $Q^2=2.45$ (Figure~\ref{dsdu_Jlab-separated}). The curves are the prediction of the full model. Note that, contrary to the $\pi^+$ production channel, the nucleon convection current contributes to the Longitudinal cross sections of the $\omega$ production channel.

\begin{figure}[hbt]
\begin{center}
\epsfig{file=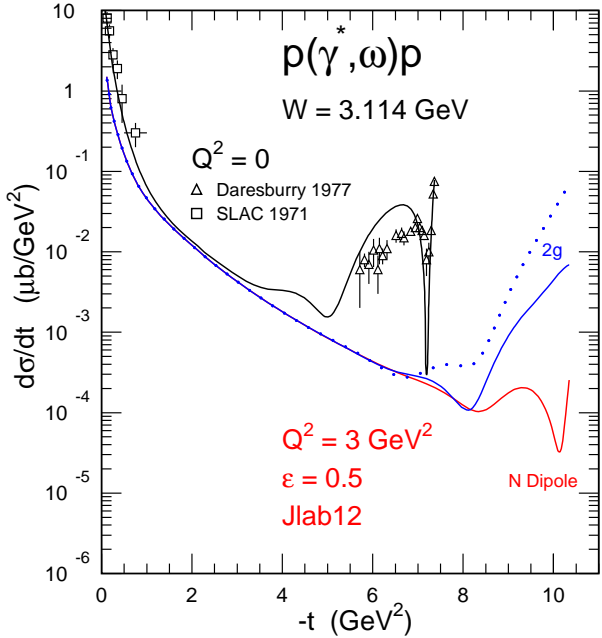,width=3.3in}

\caption[]{(Color online.) The cross section of the $p(\gamma^*,\omega)p$ reaction at $W=3.114$~GeV and $Q^2=3$~GeV$^2$ (blue curves). The black curve and experimental points  are the same as in Figure~\ref{dsdt_real}.  The meaning of the blue curves is the same as in Figure~\ref{dsdt_hermes_piplus}.}
\label{dsdt_hermes_omega}
\end{center}
\end{figure}

\begin{figure}[hbt]
\begin{center}
\epsfig{file=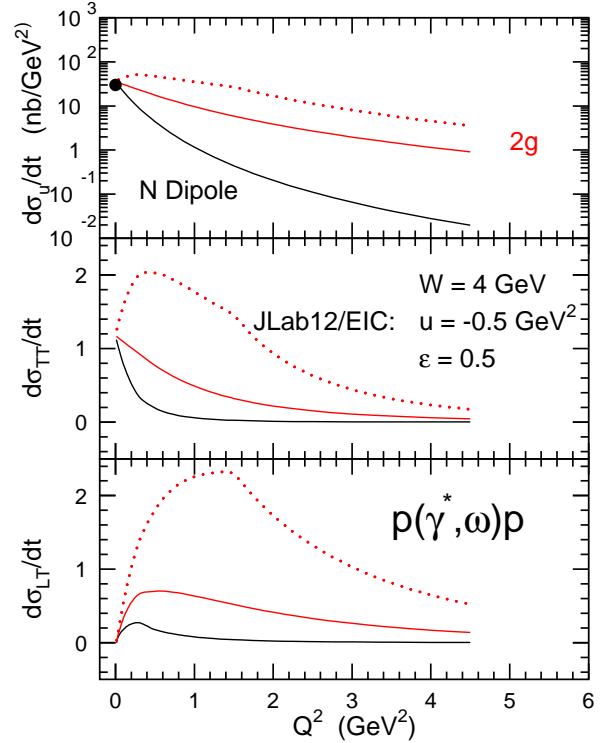,width=3.3in}

\caption[]{(Color online.) The cross section of the $p(\gamma^*,\omega)p$ reaction at $W=4$~GeV and $u=-0.5$~GeV$^2$. The black dot at the real photon point is the  experimental datum~\cite{Cl77}. The meaning of the curves is the same as in Figure~\ref{u0p5_hermes_piplus}.}
\label{u0p5_hermes_omega}
\end{center}
\end{figure}

The predictions of the model at higher energies, accessible at JLab12~\cite{Li20} and EIC~\cite{EICxx}, are given in Figures~\ref{dsdt_hermes_omega} and~\ref{u0p5_hermes_omega}. In the first, the comparison of the model with existing data at the photon point (see section~\ref{subsec:omega-real}) are shown for reference. At high virtuality $Q^2=$~3 GeV$^2$, the coupling to the $\rho^0 p$ cut (blue solid line curve) and charged $\rho N,\Delta$ cuts (blue dotted line curve) boost the backward angle cross section, well above the prediction where the nucleon canonical dipole form factor is used in both the nucleon pole and $u$-channel cut amplitudes (red solid line curve).

The second (Figure~\ref{u0p5_hermes_omega}) shows the evolution of the backward angle cross section, from the real photon point toward high photon virtualities $Q^2$. Again, the contributions of the $\rho^0 p$ $u$-channel cut (red solid line curve) and of the charged $\rho N,\Delta$ $u$-channel cuts (red dotted line curve) overwhelm  the contribution of the nucleon pole exchange (black solid line curve). A measurement at JLab12 or EIC would be worth to confirm  this conjecture which was triggered by the unexpectedly large experimental results~\cite{Li19, Pa18} recorded at lower energies (Figure~\ref{dsdt_virtual}).

\subsection{The link with TDA}

The hadronic approach that is proposed in this publication relies on the interplay between a few dominant coupled channels. It is based on the Regge poles and cuts phenomenology, and takes advantage of the known energy and momentum behavior of the elementary amplitudes. Since the experimental data basis is scarce at backward angles, it is not possible to calibrate all the rescattering integrals, as it has been done at forward angles~\cite{La10}. Therefore the rescattering integrals are parameterized by Regge cuts which encompass the driving energy and momentum dependencies. The remaining dependencies against the energy $W$ and the virtuality $Q^2$ follow those of already measured cross sections of the elementary channels which are coupled with the meson photoproduction channels.

As the energy increases, more and more coupled channels become opened, and the methods may become tedious and difficult to control. It may be more economic to rely on the direct coupling of the photon to the quarks in the nucleon. Also as the virtuality $Q^2$ increases,  the process becomes able to distinguish quarks inside the hadrons, and QCD  drives the direct coupling of the photon to these quarks.

Such a partonic approach has been proposed~\cite{Pi05, Lan11, Pi15, Pi21}. It relies on the factorization of  the non perturbative Transition Distribution Amplitudes (TDA), which describe at the quark level the transition between the nucleon and the meson, and the direct perturbative coupling of the virtual photon to the three quarks which are exchanged in the $u$-channel.  While such a factorization is justified at asymptotic energies and virtualities,  its validity at lower energies is an open question.

In fact, the two approaches must lead to similar results provided that the available energy is large enough to sum up over  the full basis of hadronic coupled channels and replace it by a quark state basis. This condition has not been fulfilled  in the JLab6 energy range, where only a few channels still dominate.  Experiments at higher energies, at Jlab12 or EIC for instance, may provide clues to address this question. 

\section{Conclusion}

The hadronic approach provides us with a unified representation of the cross sections in the real photon sector as well as the virtual photon sector.

In the real photon sector, the model leads by construction to same results as those summarized in~\cite{La19}. In the virtual photon sector, it provides us with a sensible interpretation of the surprisingly large cross sections recently recorded at backward angles in $\omega$ and $\pi$ meson electro-production, and predicts their non trivial variation with the photon virtuality $Q^2$, without destroying the good agreement at forward and intermediate angles

Since it is based on unitarity and on the known behavior of the elementary channel cross sections, this conjecture is expected to survive in a full calculation: Details may change, but one cannot escape these dominant processes.

In summary, recent data can again be understood within a complex but elegant architecture: it explains the non trivial transition between reactions induced by real and virtual photons, and relates them to other channels. The confirmation of this conjecture will greatly benefit from future measurements at JLab12 and EIC.

\acknowledgments

I acknowledge the warm scientific hospitality at JLab where recent unexpected findings triggered this work. Jefferson Science Associates operate Thomas Jefferson National Facility for the United States Department of Energy under contract DE-AC05-06OR23177.

\appendix

\section{Evaluation of $R_{\Delta}$}
\label{AA}

This appendix gathers the technical determination of the ratio $R_{\Delta}$ between the charged $\rho \Delta$ and $\rho N$ cut amplitudes in Equation~\ref{Fc_full}.

In the $\pi^0$ production channel, at lowest order, the structure of the lower part of right bottom row and left middle row diagrams is very similar~\cite{La11}:
\begin{eqnarray}
\vec{\sigma} \cdot \vec{k_{\rho}} \vec{\sigma} \cdot \vec{k_{\pi}}
&=& \vec{k_{\rho}}\cdot \vec{k_{\pi}}
+i \vec{\sigma} \cdot \vec{k_{\rho}} \times \vec{k_{\pi}}
\nonumber \\
\vec{S^{\dagger}} \cdot \vec{k_{\rho}} \vec{S} \cdot \vec{k_{\pi}}
&=& \frac{2}{3}\vec{k_{\rho}}\cdot \vec{k_{\pi}}
+i \frac{1}{3}\vec{\sigma} \cdot \vec{k_{\rho}} \times \vec{k_{\pi}}
\label{spin_amp}
\end{eqnarray}
Where $\vec{S}$ is the spin operator of the $N\rightarrow \Delta$ transition. Assuming that the scalar and vector parts contribute equally:
\begin{eqnarray}
R_{\Delta} & = &\frac{1}{2}\frac{G_{\rho }G_{\pi}}{g_{\rho }g_{\pi}}(\frac{1}{3}+\frac{1}{3})
\nonumber \\
R_{\Delta} & = & 1.5
\end{eqnarray}
where the ratio of the $\Delta$ and nucleon coupling constants is $G_{\rho }G_{\pi}/g_{\rho }g_{\pi} = 4.49$, according to~\cite{La19} and~\cite{BL77}. The last parenthesis contains the ratio of isospin coefficients: $1/3$ for each diagram involving the $\Delta$.

In the $\pi^+$ production channel, the spin structure of the $\rho^+ n$ and $\rho^- \Delta^{++}$ cut amplitudes is similar. The coupling constant are the same and the ratio between the two amplitudes depends only on the isospin coefficients:
\begin{equation}
\widetilde{R_{\Delta}} = 1
\end{equation}
However, the right part of the $\rho^+ n$ cut involves the exchange of the $\Delta$, while the right part of the  $\rho^0 p$ cut involves the exchange of the neutron. The ratio of the corresponding amplitude is not simply the ratio of the $\rho^+$ and $\rho^0$ production amplitude and the form factor, Equation~\ref{Fc_full}, should be customized:
\begin{eqnarray}
F_c(Q^2,W)&=& F^{2g}_c(Q^2)\left[1+i\;M(Q^2,W) \right.
\nonumber \\
&&\left.\frac{1}{2}\frac{1}{2}\frac{G_{\rho }G_{\pi}}{g_{\rho }g_{\pi}}
\left(1+\widetilde{R_{\Delta}}
\frac{p_{\Delta}m_{\Delta}}{pm}\right)\right]
\label{Fc_full_pi+}
\end{eqnarray}

In the kinematics considered in this work, the phase space ratio $p_{\Delta }m_{\Delta}/{pm}$ is close to unity: $0.915$ at $W=2.3$~GeV,  and closer to unity at higher energies. So, Equation~\ref{Fc_full_pi+} can be rearranged in the same form as Equation~\ref{Fc_full}, with $R_{\Delta} = 1.25$

In the calculation reported in this paper the value $R_{\Delta} = 1.5$ has been used.

In the $\omega$ production channel, both $\rho^{\pm}\Delta$ cuts (bottom part of Figure~\ref{graphs}) contain a $\Delta$ extra propagator in the $u$-channel. Under the assumption that the spin structure of the $\omega \Delta \Delta$ vertex is the same as the structure of the  $\omega NN$ vertex:
\begin{eqnarray}
R_{\Delta} & = &\frac{1}{2}\frac{G_{\rho }G_{\pi}}{g_{\rho }g_{\pi}}
\frac{G_{\omega}}{g_{\omega}}(\frac{1}{6}+\frac{1}{2})
\nonumber \\
R_{\Delta} & = & 1.9
\end{eqnarray}
where $G_{\rho }G_{\pi}/g_{\rho }g_{\pi} = 4.49$, according to~\cite{La19} and~\cite{BL77}, and assuming that the unknown ratio $G_{\omega}/g_{\omega}=1.26$. The last parenthesis contains the ratio of isospin coefficients: $1/6$ and $1/2$ for each diagram involving the $\Delta$.

The values of $R_{\Delta}$ that have been used in the calculations reported in this paper are gathered in Table~\ref{table_elastic_cut}.

\section{Virtual $\gamma$ absorption cross sections}
\label{AB}

The virtual photon absorption  cross sections are defined according to the following convention:
\begin{eqnarray}
2\pi\frac{d\sigma}{dt d\phi}&=&\frac{d}{dt}\left(\sigma_T +\epsilon \sigma_L+\epsilon cos(2\phi) \sigma_{TT}
\right. \nonumber \\
&&\left.  +\sqrt{2\epsilon(\epsilon + 1)} cos(\phi) \sigma_{LT}
\right)
\label{sigma}
\end{eqnarray}
where $\sigma_T$ stands for the Transverse cross section, $\sigma_L$ stands for the Longitudinal cross section, while $\sigma_{TT}$ and $\sigma_{LT}$ stand respectively for the Transverse-Transverse and Longitudinal-Transverse interference cross sections. The polarization of the virtual photon is $\epsilon$, while the azimuthal angle between the electron scattering plane and the hadron emission plane is $\phi$.

The unpolarized cross section  is  $\sigma_u=\sigma_T +\epsilon \sigma_L$.

More details can be found in Appendix B of~\cite{La19}.

\end{document}